\documentclass[aps,prb,twocolumn,groupedaddress,amsmath,amssymb,amsfonts,graphics]{revtex4}

\DeclareMathOperator{\imag}{Im}
\DeclareMathOperator{\sgn}{sgn}

\newcommand{\abs}[1]{\bigl\lvert #1 \bigr\rvert}
\newcommand{\bracket}[1]{\langle #1 \rangle}
\usepackage{graphicx}
\usepackage{bm}
\usepackage{bbold}
\renewcommand{\vec}[1]{\bm{#1}}

\begin{document}

\title{Ground state properties of a Zeeman-split heavy metal}

\author{K.\ S.\ D.\ Beach}
\email[]{ksdb@bu.edu}
\homepage[]{www.ksdb.name}
\affiliation{Department of Physics, Boston University}

\date{September 29, 2005}

\begin{abstract}
A Zeeman field affects the metallic heavy fermion ground state in two ways: (\emph{i}) it splits the 
spin-degerenate conduction sea, leaving spin up and spin down Fermi surfaces with different 
band curvature; (\emph{ii}) it competes with the Kondo effect and thus suppresses the mass
enhancement. Taking these two effects into account, we compute the quasiparticle effective 
mass as a function of applied field strength within hybridization mean field theory. We also 
derive an expression for the optical conductivity, which is relevant to infrared spectroscopy
measurements.
\end{abstract}

\maketitle
\section{Introduction}

Heavy fermion materials~\cite{Stewart84,Lee86} are metallic alloys of actinide or rare-earth elements
(typically U or Ce). At high temperature, their chemically active valence electrons are confined
in localized $f$ orbitals.
These $f$-electrons constitute a dense lattice of localized spins embedded in---and only
weakly interacting with---the ordinary conduction sea.
At low temperature, the $f$-electron moments become strongly coupled to the 
conduction electrons and, indirectly, to each other. Below the characteristic Kondo temperature,
a complicated many-body state emerges in which the local moments are screened
(provided that the Kondo physics overwhelms the competing RKKY interaction~\cite{Doniach77}). 
This state exhibits unconventional metallic behaviour with an effective mass tens or hundreds of 
times larger than that of a bare band electron.

The key detail is that the broad band of conduction electrons is intersected by a nearly dispersionless, highly correlated, $f$-electron band.  The heavy fermion ground state can be understood to be a 
``nearly-broken symmetry'' state~\cite{Coleman87a} in which the overlap between excitations in the two bands becomes macroscopically important (to order $1/N$, where $N \rightarrow \infty$
represents the limit of large orbital degeneracy~\cite{Zhang83,Coleman87a,Millis87a}).
This motivates treating the Kondo physics, in the guise of an interband hybridization, at the mean field level.

When the hybridization order parameter condenses, the $f$-electrons are incorporated into an enlarged Fermi sea of composite quasiparticles. Mixing between the local and itinerant degrees of freedom causes the conduction band to break into upper and lower quasiparticle bands, leaving a region of shallow dispersion (\emph{i.e.}, large effective mass) near the hybridization gap edge.

The hybridization picture has been verified in real materials by a variety of experimental methods.
Infrared (IR) spectroscopy in particular has proved to be an important experimental probe: 
IR studies of YbFe${}_4$Sb${}_{12}$ and CeRu${}_4$Sb${}_{12}$ provided the first
direct observation of the hybridization gap in heavy metals.~\cite{Dordevic01}
In addition, the complex optical conductivity and the dielectric function, obtained from reflectance measurements by Kramers-Kronig, can be used to infer the effective mass.
Such an analysis is possible because the electrodynamic
response of the system depends on the mass enhancement factor in a known way.~\cite{Coleman87b,Millis87b,Degiorgi99}

In a recent experiment, Dordevic \emph{et al}.\ carried out a detailed IR study of the heavy fermion material CeRu${}_4$Sb${}_{12}$ in high magnetic fields.~\cite{Dordevic05}
A complete theory of the electrodynamic response was lacking in this case.
Here, the applied field breaks the spin degeneracy of the Fermi sea, with the result that 
spin up and spin down quasiparticles have different effective masses.
The reflectance, however, is insensitive to the quasiparticle spin. 
Thus, the measured optical conductivity, insofar as it depends on the (now spin-dependent)
mass enhancement factor, represents an average over the two spin projections.

In this paper, we compute the mean field heavy fermion ground state in the presence of a Zeeman
field (under the assumption that the field does not induce a magnetic instability~\cite{Beach04,Milat04}). 
We find that the effective mass is modified to linear order because of Fermi surface splitting
and to quadratic order because of magnetic suppression of the Kondo singlet amplitude.
As it turns out, the optical conductivity averages over the effective mass in such a way that 
corrections first appear at second order and with a \emph{negative} coefficient;
the characteristic field scale is on the order of the Kondo energy (in magnetic units).
Such behaviour is observed experimentally in Ref.~\onlinecite{Dordevic05}.

The paper is organized as follows.
The hybridization mean field equations are derived in
Secs.~\ref{SECT:Formalism} and \ref{SECT:HFmetal}.
Their solution in zero and nonzero field is presented in Sec.~\ref{SECT:MFSolution},
followed by a discussion of 
the optical conductivity in Sec.~\ref{SECT:OptCond}.

\section{\label{SECT:Formalism}Mean Field Formalism}

\subsection{\label{SECT:HybrField}Hybridization Field}

The Kondo Lattice Model (KLM) is thought to provide an approximate description of heavy fermion materials.  
(The Periodic Anderson Model in the limit of large onsite repulsion is an equivalent
starting point.~\cite{Hewson97})
It describes a band of free conduction electrons moving in a periodic array of magnetic impurities $\hat{\vec{S}}_{\vec{r}}$. 
The only interaction is the onsite Heisenberg exchange between the electron spin density and the magnetic moment of the impurities. 
\begin{equation} \label{EQ:Hamiltonian}
\hat{H}_{\text{KLM}} = -\sum_{\vec{r}\vec{r}'} t_{\vec{r}-\vec{r}'} c_{\vec{r}}^{\dagger}c_{\vec{r}'}
    + J\sum_{\vec{r}} \hat{\vec{s}}_{\vec{r}} \cdot \hat{\vec{S}}_{\vec{r}} .
\end{equation}
Here, $t_{\vec{r}} = t_{-\vec{r}}^*$ and $J$ are the hopping and exchange integrals, respectively,
and $c^{\dagger}$ ($c$) is the creation (annihilation) operator for the
conduction electrons; $\hat{\vec{s}} = \tfrac{1}{2}c^{\dagger}\vec{\sigma}c$ describes their local moment. The impurity spins are $S=1/2$.  The model can be extended to include a Zeeman term, which consists of a magnetic field $\vec{B}$ coupled to the total magnetic moment at each site: 
\begin{equation} \label{EQ:Hamiltonian}
\hat{H} = \hat{H}_{\text{KLM}} - \vec{B} \cdot \sum_{\vec{r}} \bigl( \hat{\vec{s}}_{\vec{r}}
+ \hat{\vec{S}}_{\vec{r}} \bigr).
\end{equation}

A useful formal trick is to represent the local spins in terms of fermions, which puts the local and itinerant degrees of freedom on the same footing. Specifically, we take  $\hat{\vec{S}} = \frac{1}{2}f^{\dagger}\vec{\sigma}f$ subject to the constraint $f^{\dagger}\!f = 1$. Here, $f^{\dagger}$ ($f$) is the creation (annihilation) operator of a fictitious, dispersionless $f$ band.  The constraint suppresses all $f$-charge fluctuations and has the effect of projecting out the singlet states that are not a part of 
the Hilbert space of the original SU(2) spin.  We shall assume that it is sufficient to enforce the constraint on average and require only that $\bracket{f^{\dagger}\!f} = 1$.

The operators $\hat{\chi}^{\mu} = \tfrac{1}{\sqrt{2}}f^{\dagger}\sigma^{\mu}c$, defined in terms of the unit and Pauli matrices $\sigma^{\mu} = (\mathbb{1},\vec{\sigma})$, describe the spin degrees of freedom at each site.  They can be used to express the exchange term in the Hamiltonian as
\begin{equation} \label{EQ:decompose}
\tfrac{1}{4}c^\dagger \vec{\sigma} c \cdot f^\dagger \vec{\sigma} f
= -\tfrac{3}{4}\hat{\chi}^{0\dagger}\hat{\chi}^0+\tfrac{1}{4}
 \hat{\vec{\chi}}^{\dagger}\cdot\hat{\vec{\chi}}.
\end{equation}
$\hat{\chi}^0$ and $\hat{\vec{\chi}}$ act in the singlet and triplet channels, respectively. In the heavy fermion state, the singlet amplitude condenses $\bracket{\hat{\chi}^0} \neq 0$, leading to hybridization of the $c$ and $f$ bands. Neglecting second order fluctuations 
about $\hat{\chi}^0 = \bracket{\hat{\chi}^0}$ gives
\begin{equation}\label{EQ:nofluctuations}
\hat{\chi}^{0\dagger}\hat{\chi}^0 = \bracket{\hat{\chi}^0}^*\hat{\chi}^0
+\hat{\chi}^{0\dagger}\bracket{\hat{\chi}^0}
-\abs{\bracket{\hat{\chi}^0}}^2.
\end{equation}
The resulting mean field Hamiltonian is
\begin{equation} \label{EQ:MFHamiltonian}
\begin{split}
\hat{H}_{\text{MF}} &= -\sum_{\vec{r}\vec{r}'} c_{\vec{r}}^{\dagger}\Big[ t_{\vec{r}-\vec{r}'} + \delta_{\vec{r}\vec{r}'}\big(\mu_c + \vec{B}\cdot\vec{\sigma}/2\bigr) \Bigr]c_{\vec{r}'} \\
&\quad -\sum_{\vec{r}} f_{\vec{r}}^{\dagger}\big(\mu_f + \vec{B}\cdot\vec{\sigma}/2\bigr) f_{\vec{r}} \\
     &\quad - \sum_{\vec{r}} \bigl(V^*_{\vec{r}} f_{\vec{r}}^{\dagger}c_{\vec{r}} + V_{\vec{r}} c_{\vec{r}}^{\dagger} f_{\vec{r}} \bigr) + \frac{8}{3J}\sum_{\vec{r}} \abs{V_{\vec{r}}}^2.
\end{split}
\end{equation}
We have expressed the Hamiltonian in terms of a hybridization field
\begin{equation} \label{EQ:selfconsistentV}
V = \frac{3J}{4\sqrt{2}}\bracket{\hat{\chi}^0} = \frac{3J}{8}\bracket{f^\dagger c} 
\end{equation}
having units of energy. We have also included two chemical potentials $\mu_c$ and $\mu_f$,
which couple to the $c$- and $f$-electron densities. These allow us to control the conduction band filling and the $f$-level occupancy.

\subsection{Quasiparticle Dispersion Relation}

The minimum-energy hybridization field configuration is  translationally invariant. Thus, we can write 
$V_{\vec{r}} = V = \abs{V}e^{i\theta}$.  Moreover, although the hybridization $\langle f^\dagger c \rangle$ is in general complex, we may safely take $V$ to be real and positive. There is a U(1) gauge freedom associated with the invariance of $\hat{\vec{S}} = \frac{1}{2}f^\dagger \vec{\sigma} f$ under the phase rotation $f \mapsto e^{i\phi}f$. Accordingly, by fixing $\phi = -\theta$, we can gauge away the phase of $V$.

Equation~(\ref{EQ:MFHamiltonian}) has a particularly simple wavevector representation,
\begin{equation}
H_{\text{MF}} = \sum_{\vec{k}s} \begin{pmatrix} c^\dagger_{\vec{k}s} & f^\dagger_{\vec{k}s}\end{pmatrix}
 M_{\vec{k}s} \begin{pmatrix} c_{\vec{k}s} \\ f_{\vec{k}s}\end{pmatrix} + \frac{8NV^2}{3J}.
\end{equation}
In the summation, $\vec{k}$ ranges over all wavevectors in the Brillouin zone and
$s$ over the two fermion spin projections. $N$ denotes the number of lattice sites.
The coefficient matrix
\begin{equation} \label{EQ:Mkcoeff}
M_{\vec{k}s} = \begin{pmatrix} \epsilon_{\vec{k}} - \mu_c - sB/2
& -V \\ -V & -\mu_f - sB/2
\end{pmatrix}
\end{equation}
is a function of the free conduction-electron dispersion
$\varepsilon_{\vec{k}} = -\sum_{\vec{r}}e^{-\vec{k}\cdot \vec{r} }t_{\vec{r}}$,
the physical parameters $J$ and $B$, and the mean field
parameters $\mu_c$, $\mu_f$, and $V$.  By writing $\vec{B}\cdot\vec{\sigma}_{ss'} = B\sigma^3_{ss'} = sB\delta_{ss'}$, we have chosen to direct the applied magnetic field along the 3 axis.

It is convenient to introduce the quantities
\begin{equation} \label{EQ:mu_b}
2\mu = \mu_c + \mu_f \quad \text{and} \quad
b = \mu_c - \mu_f ,
\end{equation}
which serve as an alternate set of Lagrange multipliers (rotated 45 degrees with respect to the original set). As we shall see, $\mu$ is the energy required to remove a dressed quasiparticle from the top of the Fermi sea, whereas $b/2$ is the energy to remove a bare conduction electron. We can think of $\mu$ as the chemical potential of the fully interacting system. The parameter $b$ controls how many electrons are available to compensate each local spin.

Let us shift the diagonal entries in Eq.~\eqref{EQ:Mkcoeff} so as to void the lower right entry. The resulting matrix is
\begin{equation} \label{EQ:Mkcoeffoffset}
M_{\vec{k}s} + \bigl(\mu_f + sB/2\bigr) \mathbb{1} = \begin{pmatrix} \epsilon_{\vec{k}} - b
& -V \\ -V & 0
\end{pmatrix},
\end{equation}
which has eigenvalues
\begin{equation} \label{EQ:Ink}
I^n_{\vec{k}} = \frac{1}{2}\Bigl[ 
 \epsilon_{\vec{k}} - b + n\sqrt{( \epsilon_{\vec{k}} - b)^2
+ 4V^2} \Bigr],
\end{equation}
labelled by $n= \pm$.
In order to diagonalize Eq.~\eqref{EQ:Mkcoeff}, we construct a unitary transformation $U_{\vec{k}}$, whose columns are populated with the normalized eigenvectors of $M_{\vec{k}s}$:
\begin{equation} \label{EQ:Uunitary}
U = \begin{pmatrix}
U^{c+} & U^{c-}\\
U^{f+} & U^{f-}
\end{pmatrix}
= \begin{pmatrix}
\frac{-I^+}{\sqrt{(I^+)^2+V^2}} &
\frac{-I^-}{\sqrt{(I^-)^2+V^2}} \\
\frac{V}{\sqrt{(I^+)^2+V^2}} &
\frac{V}{\sqrt{(I^-)^2+V^2}}
\end{pmatrix}.
\end{equation}
Note that all the $\vec{k}$ dependence of $U_{\vec{k}}$ comes from $I_{\vec{k}}^n$
and that both $I^n_{\vec{k}}$ and $U_{\vec{k}}$ are independent of the magnetic field.

One can easily verify that Eq.~\eqref{EQ:Uunitary} is unitary and that it diagonalizes the coefficient matrix: $(U_{\vec{k}}^{\dagger} U_{\vec{k}})^{nn'} = \delta^{nn'}$ and $(U^{\dagger}_{\vec{k}} M_{\vec{k}s} U_{\vec{k}})^{nn'} = (E^n_{\vec{k}s}-\mu) \delta^{nn'}$, where
\begin{equation} \label{EQ:Eksn}
E^n_{\vec{k}s} = \frac{1}{2}\Bigl[ 
 \epsilon_{\vec{k}} - sB + n\sqrt{( \epsilon_{\vec{k}} - b)^2
+ 4V^2} \Bigr].
\end{equation}
When $V \neq 0$, the $c$- and $f$-electrons admix to produce the quasiparticles of the interacting system. Their dispersion is given by Eq.~\eqref{EQ:Eksn}. 
Note that hybridization splits the band structure into two disjoint pieces. The quantum number $n$ labels quasiparticles in the upper ($n=+$) and lower ($n=-$) bands.

In a Zeeman field, the Fermi surface is split into two distinct Fermi surfaces---one for spin up and one for spin down quasiparticles. For quasiparticles with spin projection $s$, the Fermi surface is defined by $E^n_{\vec{k}s} = \mu$ or, equivalently, $I^n_{\vec{k}} = \mu_{fs} \equiv \mu_f + sB/2$.

\subsection{Fixing the Mean Field Parameters}

The physics of the hybridization mean field theory depends on how the parameters $\mu_c$, $\mu_f$, and $V$ vary as a function of the Kondo coupling and the magnetic field. The optimal value of the hybridization strength is the one that minimizes the free energy density
\begin{equation} \label{EQ:freeFdensity}
\mathcal{F} 
  = \frac{8V^2}{3J}-\frac{1}{\beta N}\sum_{\vec{k}sn} \ln \Bigl[1+
     e^{-\beta (E^n_{\vec{k}s}-\mu)}\Bigr].
\end{equation}
The Lagrange multipliers are chosen to satisfy $\langle c^\dagger c\rangle = -\partial \mathcal{F}/\partial \mu_c = n_c$, where $n_c$ is the conduction band filling, and $\langle f^\dagger\!f \rangle = -\partial \mathcal{F}/\partial \mu_f = 1$. 
Expressing these conditions in terms of the rotated Lagrange multipliers of Eq.~(\ref{EQ:mu_b}), 
we find that
\begin{equation}
-\frac{\partial \mathcal{F}}{\partial \mu} = n_c + 1, \ \ 
-2\frac{\partial \mathcal{F}}{\partial b} = n_c - 1, \ \ \text{and} \ \
\frac{\partial \mathcal{F}}{\partial V} =0.
\end{equation}
Performing the free energy differentiations yields
\begin{subequations} \label{EQ:MFall}
\begin{align}
\frac{1}{N}\sum_{\vec{k}sn}f(E^n_{\vec{k}s}-\mu) &= n_c + 1 \label{EQ:MF1}\\
\frac{1}{N}\sum_{\vec{k}sn}\frac{n(\epsilon_{\vec{k}}-b)f(E^n_{\vec{k}s}-\mu)}
{\sqrt{(\epsilon_{\vec{k}}-b)^2 + 4V^2}} &= n_c - 1 \label{EQ:MF2}\\
\frac{8}{3J} + \frac{1}{N}\sum_{\vec{k}sn}
\frac{n f(E^n_{\vec{k}s}-\mu)}{\sqrt{(\epsilon_{\vec{k}}-b)^2 + 4V^2}} &= 0. \label{EQ:MF3}
\end{align}
\end{subequations}
Note that Eq.~\eqref{EQ:MF1} defines the  Luttinger volume as $n = n_c+1$, with both the $c$- and $f$-electrons counted in an enlarged Fermi sea. Equation~\eqref{EQ:MF3} is the gap equation that determines $V$.

\subsection{Thermodynamic Limit}

In pursuit of a solution to Eqs.~\eqref{EQ:MFall}, it is helpful to eliminate the $\vec{k}$ 
summations in favour of energy integrals weighted by the density of states (DOS),
\begin{equation} \label{EQ:fieldindDOS}
D(\omega) = \frac{1}{N}\sum_{\vec{k}n}\delta(\omega - I^n_{\vec{k}}).
\end{equation}
In the thermodynamic limit $(N \rightarrow \infty)$, the set of $I^n_{\vec{k}}$ values is dense, and Eq.~\eqref{EQ:fieldindDOS} is a smooth function of $\omega$. [As a convenience, we have defined the DOS as the spectrum of $I^n_{\vec{k}}$ rather than $E^n_{\vec{k}s}$, which makes the function  independent of the magnetic field. The true DOS is offset from Eq.~\eqref{EQ:fieldindDOS} by $-b/2+sB/2 = \mu_{fs}-\mu$.]
Applying the delta function identity
\begin{multline} \label{EQ:deltfuncid}
\sum_{n=\pm1 } \delta\biggl[\varepsilon_{\vec{k}}-b - 2\omega + n \sqrt{\bigl(\varepsilon_{\vec{k}}-b\bigr)^2+4V^2}\biggr]\\
= \frac{1}{2}\biggl(1 + \frac{V^2}{\omega^2}\biggr) \delta\biggl[\varepsilon_{\vec{k}}-b - \omega \biggl(1-\frac{V^2}{\omega^2}\biggr)\biggr]
\end{multline}
to Eq.~\eqref{EQ:fieldindDOS}, we can show that $D(\omega) \sim 1+V^2/\omega^2$. The correspondence between quantities in the wavevector and energy representations is summarized in Table~\ref{TAB:ksum2dos}. 

\begin{table}
\begin{center}
\begin{tabular}{ll}
\toprule
Wavevector Sum \hspace{0.3cm} & DOS Integral\\ \colrule
$I^n_{\vec{k}}$ & $\omega$\\ 
$I^{-n}_{\vec{k}}$ & $ -V^2/\omega$\\ 
$\epsilon_{\vec{k}}-b$ & $(\omega^2-V^2)/\omega$\\
$n$ & $\sgn(\omega)$\\
$\sum_{\vec{k}n}$ & $\int d\omega\, D(\omega)$ \vspace{0.1cm} \\ 
\botrule
\end{tabular}
\end{center}
\caption{\label{TAB:ksum2dos} A wavevector sum is transformed to a density of states integral by making the substitutions listed in this conversion chart. In addition
$E^n_{\vec{k}s} = \mu$ corresponds to $\omega = \mu_{fs}$.
}
\end{table}

\begin{figure}
\includegraphics{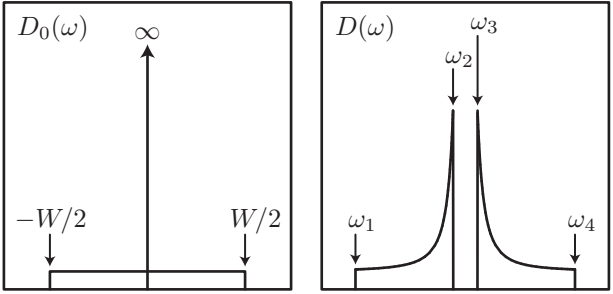}
\caption{\label{FIG:gapomegai}The total density of states for the noninteracting system $D_0(\omega)$ consists of an $f$-level delta function peak superimposed on a conduction band of width $W$. When $V\neq 0$, hybridization breaks the spectral weight into lower ($\omega_1 < \omega < \omega_2$) and upper ($\omega_3 < \omega < \omega_4$) bands.}
\end{figure}

The most important feature of the DOS is that it develops a band gap as $V$ increases from zero. The DOS of the noninteracting conduction electrons can be written as the product of a line-shape function $g(\omega)$ and a heaviside function, which ensures that the density of states vanishes outside the band:
\begin{equation} \label{EQ:DOSc0}
D^{c}_0(\omega) = \frac{1}{N}\sum_{\vec{k}}\delta(\omega - \epsilon_{\vec{k}}) = g(\omega)\theta (W^2 - 4\omega^2).
\end{equation}
In the interacting system, the conduction-electron DOS has the same basic form,
\begin{equation} \label{EQ:DOSc}
D^c(\omega) = \frac{1}{N}\sum_{\vec{k}n}\lvert U_{\vec{k}}^{cn}
\rvert^2\delta(\omega - I^n_{\vec{k}}) = g(\lambda(\omega))\theta (W^2 - 4\lambda(\omega)^2),
\end{equation}
but its energy scale is renormalized by the function $\lambda(\omega) = \omega(1-V^2/\omega^2)+b$, which comes from the delta function on the right-hand side of Eq.~\eqref{EQ:deltfuncid}. As a result, the argument of the heaviside function in Eq.~\eqref{EQ:DOSc} is a fourth degree polynomial whose roots
\begin{subequations} \label{EQ:fourroots}
\begin{align} 
\omega_1 &= -\sqrt{\bigl(W/4+b/2\bigr)^2+V^2} -W/4 -b/2 \label{EQ:root1} \\
\omega_2 &= -\sqrt{\bigl(W/4-b/2\bigr)^2+V^2} +W/4 -b/2 \label{EQ:root2} \\
\omega_3 &= +\sqrt{\bigl(W/4+b/2\bigr)^2+V^2} -W/4 -b/2 \label{EQ:root3} \\
\omega_4 &= +\sqrt{\bigl(W/4-b/2\bigr)^2+V^2} +W/4 -b/2 \label{EQ:root4}
\end{align}
\end{subequations}
delineate the band edges:
\begin{equation} \label{EQ:bandedges}
\theta(W^2 - 4\lambda(\omega)^2) = \sum_{i=1}^4 (-1)^{i+1}\theta(\omega-\omega_i).
\end{equation}
Spectral weight exists only at energies in a lower band from $\omega_1$ to $\omega_2$ and in an upper band from $\omega_3$ to $\omega_4$. The two bands are separated by a gap of width $2\Delta = \omega_3 - \omega_2$. See Fig.~\ref{FIG:gapomegai}.

A similar analysis shows that the DOS for the $f$-electrons differs from $D^c(\omega)$ by a factor of $V^2/\omega^2$. Hence, 
$D(\omega) = D^c(\omega) + D^f(\omega)$, 
the total DOS, is equal to
\begin{equation} \label{EQ:DOStotexact}
D(\omega) = \biggl(1 + \frac{V^2}{\omega^2}\biggr) g(\lambda(\omega)) \sum_{i=1}^4 (-1)^{i+1}\theta(\omega-\omega_i).
\end{equation}
Using the conversion chart in Table~\ref{TAB:ksum2dos}, we can re-express 
Eq.~\eqref{EQ:freeFdensity} as
\begin{equation}
\mathcal{F} =
     \frac{8V^2}{3J}-\frac{1}{\beta}\sum_s\int \!d\omega\,D(\omega)\ln \Bigl[1+
     e^{-\beta(\omega-\mu_{fs})}\Bigr].
\end{equation}
Hence, the constituent equations of the mean field theory [\emph{viz.}, Eqs.~\eqref{EQ:MFall}] can be written compactly as
\begin{equation} \label{EQ:consituenteqs}
\sum_s \int d\omega\,f(\omega-\mu_{fs})
\left\{\begin{matrix} D^c(\omega)\\D^f(\omega)\\ \frac{-3J}{8\omega}
D^c(\omega)\end{matrix}\right\}
= \left\{\begin{matrix} n_c \\ 1 \\ 1 \end{matrix}\right\}.
\end{equation}

\section{\label{SECT:HFmetal}Heavy Fermion Metal}

\subsection{Characterizing the Ground State}

\begin{figure}
\includegraphics{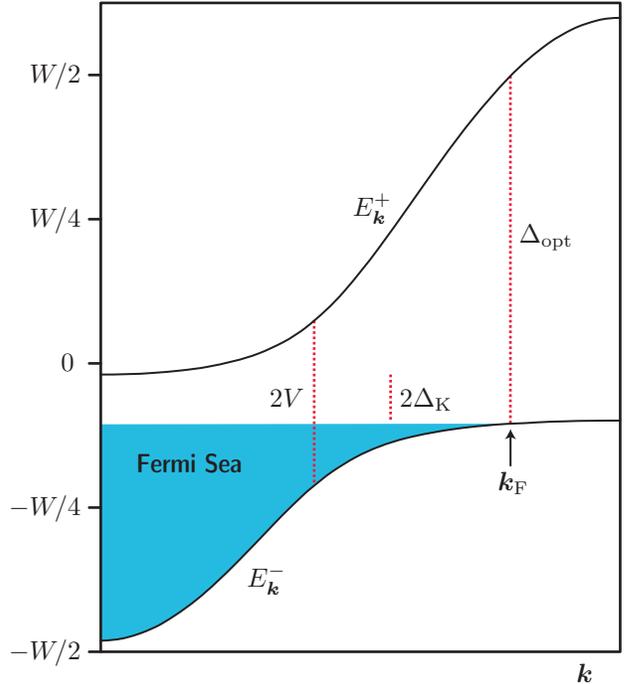}
\caption{\label{FIG:HeavyMetal} 
The hybridized band structure is plotted in zero field for conduction band filling $n_c = 3/4$
and Kondo energy $\Delta_{\text{K}} = W/25$. 
The heavy metal has a partially filled lower band $E^-_{\vec{k}}$
with quasiparticle states occupied up to the Fermi wavevector $\vec{k}_{\text{F}}$.
Several important energy scales are marked with (red) dotted lines. 
}
\end{figure}

The key feature of the heavy fermion state is the hybridization gap. It generates a region of very shallow dispersion near the gap edge, which is responsible for the large effective mass of the quasiparticles.  For concreteness, let us suppose that the lower band is filled to some point \emph{below} the hybridization gap. Then the relevant dispersion relation is that of the lower band. This situation is depicted in Fig.~\ref{FIG:HeavyMetal}. The heavy fermion state is metallic and possesses a well-defined Fermi surface given by the set of $\vec{k}$ points satisfying $E^-_{\vec{k}} = \mu$. 

The effective mass $m^*$ of the quasiparticle excitations is a function of the band curvature.
It is related to the noninteracting band mass by the variation 
$m^*\delta E^-_{\vec{k}s} = m^*\delta I^-_{\vec{k}} = m\delta \epsilon_{\vec{k}}$,
averaged over all points on the Fermi surface. 
Hence, the mass enhancement factor is given by $(m^*/m) = (\partial I^n_{\vec{k}}/\partial \varepsilon_{\vec{k}})^{-1}_{\text{F}}$ and, via Table~\ref{TAB:ksum2dos},
\begin{equation} \label{EQ:effmassform}
\biggl(\frac{m^*}{m}\!\biggr)_{\!\!s}
 = \frac{\partial}{\partial \omega}\biggl(\frac{\omega^2-V^2}{\omega}
 \biggr)_{\text{F}}
= 1 + \frac{V^2}{\omega^2}\biggr\rvert_{\omega = \mu_{fs}}.
\end{equation}

The direct energy splitting between the bands can be written as
\begin{equation} \label{EQ:upperlowersplit}
E^+_{\vec{k}s} - E^-_{\vec{k}s} = \frac{1}{2}\sum_n n\bigl(I^n_{\vec{k}} - I^{-n}_{\vec{k}}\bigr).
\end{equation}
Since optical experiments probe quasiparticles at the top of the Fermi sea by promoting them 
to the upper band with negligible momentum transfer and 
without inducing spin flips ($d\vec{k} = 0$, $s$ fixed),
the optical gap is determined by evaluating Eq.~\eqref{EQ:upperlowersplit} at the Fermi level.
Again, via Table~\ref{TAB:ksum2dos}, 
\begin{equation} \label{EQ:optgap}
\bigl(\Delta_{\text{opt}}\bigr)_{\!s} = \frac{1}{2}\sgn(\omega)\biggl(\omega + \frac{V^2}{\omega}\biggr)\biggr\rvert_{\omega = \mu_{fs}}.
\end{equation}

Note that Eqs.~\eqref{EQ:effmassform} and \eqref{EQ:optgap}
are equal up to an overall factor proportional to the 
$f$-level energy:
\begin{equation} \label{EQ:specialratio}
\biggl(\frac{\Delta_{\text{opt}}}{m^*/m}\biggr)_{\!s} = \frac{1}{2}\lvert \mu_{fs} \rvert.
\end{equation}

\subsection{Mean Field Equations in Detail}

Let us represent the bare conduction band filling by $n_c = 1-x$. (The volume enclosed by the
enlarged Fermi surface is $n_c + n_f = 2-x$.) We can think of $x$ as the density of holes in the lower band doping the system away from the half-filled Kondo insulator state.
Further, since Van Hove singularities do not play an important role here, let us assume that the density of levels in Eq.~\eqref{EQ:DOStotexact} is flat and replace the line-shape function by its average value $g \rightarrow 1/W$. The total DOS is then
\begin{equation}
D(\omega) = \frac{1}{W}\biggl(1 + \frac{V^2}{\omega^2}\biggr) \sum_{i=1}^4 (-1)^{i+1}\theta(\omega-\omega_i).
\end{equation}

Expanding Eqs.~\eqref{EQ:fourroots} in $V^2/(W\pm b)^2 \ll 1$, we find that the bottom of the band is given by
\begin{equation} \label{EQ:bottomband}
\omega_1 = -\frac{W}{2} - b -\frac{4V^2}{2(W+2b)} + \frac{(4V^2)^2}{2(W+2b)^3}
\end{equation}
and the hybridization gap edges by
\begin{subequations} \label{EQ:gapedges}
\begin{align}
\omega_2 &= -\frac{4V^2}{2(W-2b)} + \frac{(4V^2)^2}{2(W-2b)^3}, \label{EQ:lowergapedge}\\
\omega_3 &= +\frac{4V^2}{2(W+2b)} - \frac{(4V^2)^2}{2(W+2b)^3}.
\end{align}
\end{subequations}
Defining the gap width $2\Delta = \omega_3 - \omega_2$, we can invert Eqs.~\eqref{EQ:gapedges} to solve for the hybridization strength:
\begin{equation} \label{EQ:hybrstrength}
4V^2 = 2\Delta\biggl[ \bigl(W+2\Delta\bigr) - \bigl(W-6\Delta\bigr)\biggl(\frac{2b}{W}\biggr)^2\biggr].
\end{equation}

Since $b$ controls the difference between the $c$- and $f$-electron occupation, it is proportional to the number of holes in the lower band. We can express $b = -Wx/2 + \epsilon$ as the noninteracting ($J=0$) result plus a correction $\epsilon \sim \Delta$. Equations~\eqref{EQ:bottomband}, \eqref{EQ:lowergapedge}, and \eqref{EQ:hybrstrength} become
\begin{equation} \label{EQ:minus2omega1}
-2\omega_1 = W\bigl(1-x\bigr) + 2\epsilon + 2\Delta\bigl(1+x\bigr) - \frac{4\Delta\epsilon}{W} - \frac{(2\Delta)^2}{W}2x,
\end{equation}
\begin{equation} \label{EQ:minus2omega2}
-2\omega_2 = 2\Delta\bigl(1-x\bigr) + \frac{4\Delta\epsilon}{W} + \frac{(2\Delta)^2}{W}2x,
\end{equation}
and
\begin{equation} \label{EQ:4V2}
4V^2 = 2\Delta\Bigl[ W\bigl(1-x^2) + 4x\epsilon + 2\Delta\bigl(1+3x^2\bigr)\Bigr].
\end{equation}

From here, it is straightforward to sketch out how the solution to the mean field
equations is obtained. The zero-temperature $c$- and $f$-electron occupation are computed 
by integrating $D^c(\omega)$ and $D^f(\omega)$ from $\omega_1 $ up to $\mu_{fs}$. 
So long as $\mu_{fs} < \omega_2$, we can write
\begin{equation} \label{EQ:cdagceq1x}
\langle c^\dagger c \rangle = \frac{1}{W}\sum_s \int_{\omega_1}^{\mu_{fs}} \!d\omega= 
\frac{2}{W}\bigl(\mu_f - \omega_1\bigr)
\end{equation}
and
\begin{equation} \label{EQ:fdagfeq1}
\langle f^\dagger f \rangle = \frac{V^2}{W}\sum_s\int_{\omega_1}^{\mu_{fs}} \! \frac{d\omega}{\omega^2} 
= \frac{4V^2}{W} \frac{2\mu_f W(1-x)-B^2}{2\omega_1(4\mu_f^2-B^2)}.
\end{equation}
The first condition in Eq.~\eqref{EQ:consituenteqs}, $\langle c^\dagger c \rangle = 1-x$,
 implies that $2\mu_f - 2\omega_1 = W(1-x)$, which defines $-\mu_f \sim \epsilon + \Delta(1+x)$ via
Eq.~\eqref{EQ:minus2omega1}. The second condition, 
 $\langle f^\dagger f \rangle = 1$, fixes the value of $\epsilon$
in terms of the variable $\Delta$ (the only remaining unknown)
and the constants $x$ and $W$.  
Finally, the third condition,
\begin{equation} \label{EQ:gapequationB}
1 = -\frac{3J}{8W} \sum_s \int_{\omega_1}^{\mu_{fs}}\!\frac{d\omega}{\omega} 
= -\frac{3J}{4W}  \ln \biggl( \frac{\sqrt{ 4\mu_f^2-B^2}} {W(1-x) - 2\mu_f} \biggr),
\end{equation}
closes the system of equations. In the next section, we carry out these 
steps explicitly for the $B=0$ and $B\neq 0$ cases.

\section{\label{SECT:MFSolution}Mean Field Solution}

\subsection{In Zero Applied Field}

When $B=0$, the requirement that Eq.~\eqref{EQ:fdagfeq1} equal unity 
reduces to
\begin{equation}
4V^2\big(1-x\bigr) = 4\mu_f\omega_1 = -2\omega_1\bigl[ -2\omega_1
-W(1-x)\bigr].
\end{equation}
Substitution of Eqs.~\eqref{EQ:minus2omega1} and \eqref{EQ:4V2}
then allows us to solve for $\epsilon$ as a function of the hybridization gap:
\begin{equation} \label{EQ:epsilonB0}
2\epsilon = -2\Delta x\bigl(1+x\bigr) +\frac{(2\Delta)^2}{W}x^2\bigl(1-x).
\end{equation}
\begin{widetext}
Several results follow immediately. 
The energies of the bottom and top of the lower quasiparticle band are
\begin{align} \label{EQ:omega1funcDelta}
-2\omega_1 &= W\bigl(1-x\bigr) + 2\Delta\bigl(1-x^2\bigr) - \frac{(2\Delta)^2}{W} x\bigl(1-x\bigr)^2,\\
-2\omega_2 &= 2\Delta\bigl(1-x\bigr) + \frac{(2\Delta)^2}{W} x\bigl(1-x\bigr).
\intertext{The Lagrange multipliers are}
\mu_c &= -\Delta - \frac{x}{2} \biggl[ W + 2\Delta - \frac{(2\Delta)^2}{W}\bigl(1-x\bigr)\biggr],\\
\label{EQ:muffuncDelta}
\mu_f &= -\Delta + \frac{x}{2} \biggl[ 2\Delta x + \frac{(2\Delta)^2}{W}\bigl(1-x\bigr)^2\biggr],
\intertext{or, alternatively,}
\mu &= -\Delta - \frac{x}{4} \biggl[ W + 2\Delta\bigl(1-x\bigr) - \frac{(2\Delta)^2}{W}\bigl(1-x\bigr)\bigl(2-x\bigr)\biggr],\\
b &= -\frac{x}{2} \biggl[ W + 2\Delta\bigl(1+x\bigr) - \frac{(2\Delta)^2}{W}x\bigl(1-x\bigr)\biggr].
\intertext{The hybridization energy is}
\label{EQ:V2funcDelta}
4V^2 &= 2\Delta\bigl(1-x\bigr)\Bigl[ W\bigl(1+x\bigr) + 2\Delta\bigl(1 + x + 2x^2 \bigr) \Bigr].
\end{align}
\end{widetext}

As a consistency check, we verify that the assumptions made during the derivation
hold true: the $f$-level chemical potential, $\mu_f = -\Delta (1-x^2)$, does indeed
sit below the top of the lower band, $\omega_2 = -\Delta(1-x)$;
it is also true that $b \sim x$ and $\epsilon \sim \Delta$.

The value of $\Delta$ is obtained by substituting Eq.~\eqref{EQ:muffuncDelta} and $B=0$ into Eq.~\eqref{EQ:gapequationB}:
\begin{equation}
1 = -\frac{3J}{4W}  \ln \biggl( \frac{ 2\Delta(1-x^2) + \cdots} {W(1-x) + 2\Delta(1-x^2) + \cdots} \biggr).
\end{equation}
The solution of this gap equation is the Kondo energy,
\begin{equation} \label{EQ:KondoenergyHF}
2\Delta_{\text{K}} = \frac{W}{1+x}e^{-4W/3J} + \frac{(1+3x)W}{(1+x)^3}e^{-8W/3J}.
\end{equation}
For realistic values of the physical parameters, the bare exchange 
coupling is smaller than the bandwidth. 
Thus $\alpha = e^{-4W/3J} \ll 1$, which implies that
$2\Delta_{\text{K}}/W$,
\begin{equation}
\frac{-2\mu_f}{W} = \frac{\alpha(1-x)}{1-\alpha}, \text{\ \ and\ \ }
\frac{4V^2}{W^2} = \alpha\bigl(1+2\alpha\bigr)\bigl(1-x\bigr)
\end{equation}
are small parameters. The expansion leading to Eqs.~\eqref{EQ:bottomband} and \eqref{EQ:gapedges},
however, requires the stronger condition that
\begin{equation}
\frac{4V^2}{(W-2b)^2} \sim \frac{\alpha}{1-x}
\end{equation}
be small. The hybridization mean field theory is not appropriate
when $n_c = 1-x \lesssim \alpha$, the
so-called ``exhaustion limit'' of Nozi\`{e}res.~\cite{Nozieres}

Most important, small $\alpha$ guarantees that the mass enhancement factor is a large number:
\begin{equation}
\frac{m^*}{m} = \frac{1}{\alpha(1-x)} + 1.
\end{equation}
We emphasize again that the enhancement is a consequence of the small hybridization
gap and the very shallow quasiparticle dispersion at the top of the Fermi sea, as
depicted in Fig.~\ref{FIG:HeavyMetal}. 
The energy scale of the optical gap, also shown in the figure, is the conduction bandwidth. 
$\alpha$ appears only as a subleading contribution:
\begin{equation}
\Delta_{\text{opt}} = \frac{W}{2}\Bigl[ 1 + \alpha\bigl(2-x\bigr)\Bigr].
\end{equation}

\subsection{In Nonzero Applied Field}

\begin{figure}
\includegraphics{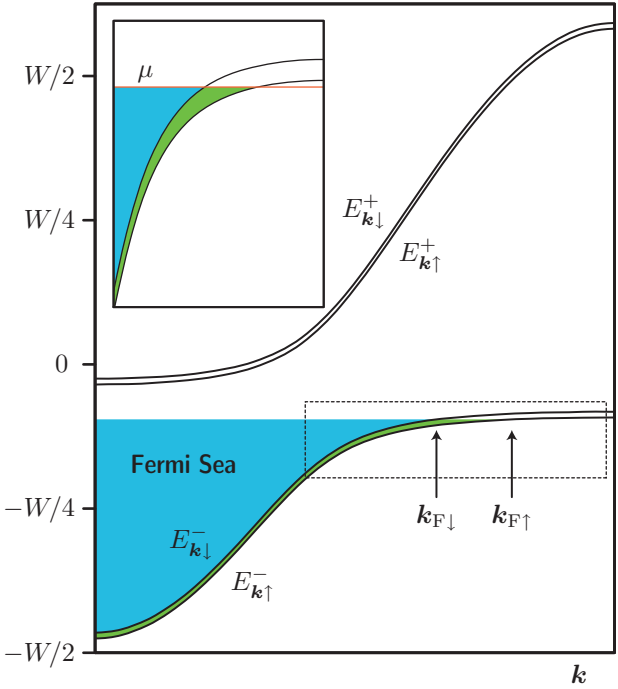}
\caption{\label{FIG:SplitHeavyMetal} 
The plot shows the band structure of Fig.~\ref{FIG:HeavyMetal} under the influence of an applied
Zeeman field. The quasiparticle bands $E^n_{\vec{k}s}$ are spin split, and the Kondo energy
$\Delta_{\text{K}}$ is somewhat reduced from its zero-field value. The inset (top-left) provides a magnified view of the (dotted) region near the lower hybridization gap edge. The difference
in band curvature at $\vec{k}_{\text{F}\uparrow}$ and $\vec{k}_{\text{F}\downarrow}$ is clearly
visible.
}
\end{figure}

When $B \neq 0$, the quasiparticles acquire a net magnetization,
$\mathcal{M} = \tfrac{1}{2}(\langle c^\dagger \sigma^3 c \rangle
+ \langle f^\dagger \sigma^3 f \rangle)$.
Since
\begin{equation} \label{EQ:csig3c}
\langle c^\dagger \sigma^3 c \rangle = \frac{1}{W}\sum_s s \int_{\omega_1}^{\mu_{fs}}\! d\omega = \frac{B}{W}
\end{equation}
and
\begin{equation} \label{EQ:fsig3f}
\langle f^\dagger \sigma^3 f \rangle = \frac{V^2}{W}\sum_s s\int_{\omega_1}^{\mu_{fs}}\! \frac{d\omega}{\omega^2} = \frac{4V^2}{W}\frac{B}{4\mu_f^2-B^2},
\end{equation}
the spin susceptibility $\chi = \partial \mathcal{M}/\partial B\rvert_{B=0}$
is Pauli-like, but enhanced by a factor of $1/\alpha$ with respect to that
of the bare conduction electrons [cf. Eq.~\eqref{EQ:effmassform}]:
\begin{equation}
\chi = \frac{1}{2W}\biggl(1 + \frac{V^2}{\mu_f^2}\biggr) = 
\chi_c \biggl(\frac{m^*}{m}\biggr)_{\!0}.
\end{equation}

In terms of the solution outlined in the previous section,
the effect of the applied field appears as a modification to Eq. \eqref{EQ:epsilonB0}.
Imposing $\langle c^{\dagger}c \rangle = 1-x$ and $\langle f^{\dagger}f \rangle = 1$
by way of Eqs.~\eqref{EQ:cdagceq1x} and \eqref{EQ:fdagfeq1} yields
\begin{multline} \label{EQ:epsilonB}
2\epsilon = \sqrt{\Delta^2(1-x^2)^2 + B^2} - \Delta(1+x)^2 \\
+\frac{(2\Delta)^2}{W}x^2\bigl(1-x) + \frac{B^2}{W}\frac{x^2+4x+1}{(1+x)^2(1-x)}.
\end{multline}
This result is correct to up to terms of order $B^4/W^3$, which are negligible
in the regime where the applied field is comparable to the Kondo energy
($B \lesssim \Delta \ll W$). To second order in $B/\Delta$, Eq.~\eqref{EQ:epsilonB}
behaves as $2\epsilon = (2\epsilon)_{B=0} + B^2/2\Delta(1-x^2)$, and thus
induces field-dependent corrections to Eqs.~\eqref{EQ:omega1funcDelta}--\eqref{EQ:V2funcDelta}
in the obvious way. In particular, we have
\begin{equation} \label{EQ:muffromepsilon}
-2\mu_f = 2\Delta(1-x^2) + \frac{B^2}{2\Delta(1-x^2)} + \cdots
\end{equation}

Once again, the system of equations is closed by appealing to Eq.~\eqref{EQ:gapequationB},
which in nonzero field admits the series solution
\begin{equation}
-2\mu_f = \frac{\alpha W(1-x)}{1-\alpha} + \frac{B^2}{2\alpha W(1-x)} -
\frac{(1-\alpha^2)B^4}{8\alpha W(1-x)}.
\end{equation}
To leading order in $\alpha$, comparison with Eq.~\eqref{EQ:muffromepsilon}
gives
\begin{equation}
2\Delta_{\text{K}} = \frac{\alpha W}{1+x} \biggl[1 - \frac{1}{2}\biggl(\frac{B}{B_0}\biggr)^2 \biggr].
\end{equation}
Here, we have introduced the characteristic field strength $B_0 = \mu_f\rvert_{B=0}
= \alpha W \bigl(1-x\bigr)$. Note that the \emph{enhancement} of the $f$-level chemical
potential is accompanied by a \emph{suppression} of the Kondo energy.
This is a consequence of the Zeeman field's favouring triplet over Kondo singlet pairing.

The corresponding results for the effective mass and optical gap are
\begin{equation} \label{EQ:effmassB}
\biggl(\frac{m^*}{m}\biggr)_{\!s} = \biggl(\frac{m^*}{m}\biggr)_{\!0}  \biggl[ 1 + 2s\biggl(\frac{B}{B_0}\biggr) + \frac{3}{2} \biggl(\frac{B}{B_0}\biggr)^2 \biggr]
\end{equation}
and
\begin{equation} \label{EQ:optgapB}
\bigl(\Delta_{\text{opt}}\bigr)_{s} = \bigl(\Delta_{\text{opt}}\bigr)_{0}  \biggl[ 1 + s\biggl(\frac{B}{B_0}\biggr) - \frac{x\alpha}{2} \biggl(\frac{B}{B_0}\biggr)^2 \biggr].
\end{equation}
In Eq.~\eqref{EQ:effmassB}, the linear correction arises from the $\mu_f + sB/2$
denominator in Eq.~\eqref{EQ:effmassform} and can be directly attributed to the spin
splitting of the Fermi sea. The quadratic correction incorporates both
Fermi-sea effects appearing at second order and Kondo energy suppression appearing
indirectly though the $V^2$ numerator. 
Equations \eqref{EQ:effmassB} and \eqref{EQ:optgapB} 
satisfy the ratio rule given in Eq.~\eqref{EQ:specialratio}.

\section{\label{SECT:OptCond}Optical Conductivity}

The total current density in the presence of a vector potential $A$
follows from the Kubo formula in the usual way:
\begin{multline}
J^{\alpha}(\vec{q},\omega) = -\frac{e^2}{c}\frac{1}{N}\sum_{\vec{k}sn} \biggl(\frac{\partial E^n_{\vec{k}s}}{\partial \epsilon_{\vec{k}}}\biggr)^2\bigl(v^{\alpha}_{\vec{k}}\bigr)^2 \delta^{\alpha\beta} 
\\
\times f'(E^n_{\vec{k}s}-\mu)
\frac{\omega}{\omega-\Bigl(\frac{\partial E^n_{\vec{k}s}}{\partial \epsilon_{\vec{k}}}\Bigr)
\vec{v}_{\vec{k}} \cdot \vec{q} + i \Gamma} A^{\beta}(\vec{q},\omega).
\end{multline}
Here, $v^{\alpha}_{\vec{k}} = \partial \epsilon_{\vec{k}}/\partial k^{\alpha}$ is the
velocity of the bare band electrons and $1/\Gamma$ is the quasiparticle lifetime.
The current density is related to the conductivity by Ohm's law,
$J = \sigma E = \sigma i\omega A/c$. Hence,
\begin{equation}
\sigma^{\alpha\beta}(\vec{q}\rightarrow 0,\omega) = \sigma(\omega)\delta^{\alpha\beta}
= \lim_{\vec{q}\rightarrow 0} \frac{c}{i\omega} \frac{\delta J^{\alpha}(\vec{q},\omega)}{\delta A^{\beta}(\vec{q},\omega)}.
\end{equation}

In the limit of zero temperature,
$f'(x) \to -\delta(x)$ restricts the $\vec{k}$ summation
to points on the Fermi surface. This allow us to pull one  
factor of $(\partial E^n_{\vec{k}s}/\partial \epsilon_{\vec{k}})_F = m/m^*$ outside
the sum. The remaining terms can be computed using integration by parts, noting that
$\partial^2 \epsilon_{\vec{k}}/\partial k^2 = 1/m$ and
$\frac{1}{N}\sum_{\vec{k}n} f(E^n_{\vec{k}s}-\mu) = n_s$.
The resulting expression for the conductivity is
\begin{equation}
\sigma = \frac{e^2}{m} \sum_s \frac{n_s}{\Gamma - i\omega (m^*/m)_s},
\end{equation}
and when $\Gamma \ll \omega$,
\begin{equation} \label{EQ:conductivitynoGamma}
\sigma = \frac{e^2}{-i \omega m} \biggl[ \frac{n_{\uparrow}(m^*/m)_{\downarrow}+n_{\downarrow}(m^*/m)_{\uparrow}}{(m^*/m)_{\uparrow}(m^*/m)_{\downarrow}}\biggr].
\end{equation}

The difference $2\mathcal{M} = n_{\uparrow} - n_{\downarrow}$ between spin up
and spin down quasiparticle occupation is obtained by
summing Eqs.~\eqref{EQ:csig3c} and \eqref{EQ:fsig3f}. Up to terms
of order $(B/B_0)^3$,
\begin{equation}
n_{\uparrow} - n_{\downarrow} = \biggl[ 1 + \frac{4V^2}{4\mu_f^2-B^2}\biggr]\frac{B}{W} = 
\frac{B}{B_0}.
\end{equation}
In other words, $n_{\uparrow} = \tfrac{1}{2}(n+B/B_0)$ and $n_{\uparrow} = \tfrac{1}{2}(n-B/B_0)$, where $n = n_{\uparrow} + n_{\downarrow} = 2-x$ is the total Luttinger volume.
Putting these expressions and 
Eq.~\eqref{EQ:effmassB} into Eq.~\eqref{EQ:conductivitynoGamma} gives
\begin{equation}
-\frac{1}{\omega} \imag \frac{1}{\sigma} = \frac{m}{e^2 n}\biggl(\frac{m^*}{m}\biggr)_{\!0}
\frac{1 - (B/B_0)^2}{1 + (3/2 - 2/n)(B/B_0)^2}.
\end{equation}

At large frequencies, the conductivity is related to the plasma frequency by
\begin{equation}
-\frac{1}{\omega} \imag \frac{1}{\sigma} \xrightarrow{\omega \rightarrow \infty} \frac{m}{e^2n} = \frac{4\pi}{\omega^2_{\text{P}}}.
\end{equation}
Thus, the expression most directly relevant to optical measurements is
\begin{equation}
-\frac{\omega^2_{\text{P}}}{4\pi\omega}\imag \frac{1}{\sigma} = \biggl(\frac{m^*}{m}\biggr)\biggl[  1 -\biggl(\frac{5}{2} - \frac{2}{n}\biggr)\biggl(\frac{B}{B_0}\biggr)^2\biggr].
\end{equation}
The coefficient $-(5/2 - 2/n) = -[3/2 + x/(2-x)]$ is negative over the
full range of band fillings.

\section{Conclusions}

The hybridization mean field theory is appropriate for intermediate
values of the exchange coupling:
to fend off competing magnetic states, $J$ must be larger than the energy scale associated with 
spontaneous ordering of the $f$-electron moments; to make contact with 
realistic metals, $J$ should also be less than the conduction bandwidth. 
In this regime, the Kondo and hybridization energies
obey $\Delta_{\text{K}}/W  \sim V^2/W^2 \sim \alpha$,
where $\alpha = e^{-4W/3J}$ is a small parameter.
The range of valid band fillings, corresponding to
$ 0 < x \lesssim 1 - \alpha$,
lies between the Kondo-insulator and exhaustion limits.

The hybridization picture captures the essential features of a heavy metal.
It allows us to understand how quantities such as the effective mass 
and magnetic susceptibility are renormalized by a common enhancement factor $1/\alpha$.
It also explains the existence of the upper quasiparticle band and clarifies the role of the 
hybridization gap. Most important for our purposes, the mean field theory provides a simple 
framework within which to compute transport  and electrodynamic properties.

In this paper, we have extended the usual heavy fermion picture to the case
of an applied Zeeman field. The spin splitting of the Fermi surface
forces us to distinguish between the spin up and spin down quasiparticles.
We have shown that their effective masses differ to linear order in $B/B_0$.
Further contributions appear at second order because of a reduction in
the Kondo energy. Accounting for these two effects, we have arrived at a
specific prediction for the behaviour of 
the optical conductivity.

Our treatment makes the assumption ($\mu_{fs} < \omega_2$)
that the applied field is suitably small with respect to the Kondo energy:
$\lvert B \rvert < 2\Delta_{\text{K}}x(1-x)$.
From an experimental point of view, this may still be a large field.
The Kondo energies of typical heavy fermion materials are
in the range of 40--100~K. 
Hence, the predictions in this paper may be valid for fields
as large as 20--30~T.

The author would like to thank 
Dimitri Basov and Sasha Dordevic
for many helpful discussions.

\end{document}